\documentclass[superscriptaddress, noeprint, amsmath,amssymb,aps,twocolumn]{revtex4-2}
\usepackage{graphicx}
\usepackage{dcolumn}
\usepackage{xcolor}
\usepackage{bm}
\usepackage{hyperref}
\usepackage{upgreek}
\usepackage{fixme}

\usepackage{verbatim}

\newcommand{\kb} {\ensuremath{k_{\mathrm{B}}}}

\hypersetup{bookmarksnumbered=true,	unicode=false,	pdfstartview={FitH}, pdfnewwindow=true, colorlinks=true,linkcolor=blue, citecolor=blue, filecolor=blue, urlcolor=blue}

\newcommand{\figref}[2]{Fig. \ref{#1}{\color{blue}#2}}%

\begin{document}

\title{Evidence for nodal superconductivity in infinite-layer nickelates}

\author{Shannon P. Harvey}
\thanks{These authors contributed equally}%
\affiliation{Stanford Institute for Materials and Energy Sciences, SLAC National Accelerator Laboratory,  Menlo Park, CA USA}
\affiliation{Department of Applied Physics, Stanford University, Stanford, CA, USA}

\author{Bai Yang Wang}
\thanks{These authors contributed equally}
\affiliation{Stanford Institute for Materials and Energy Sciences, SLAC National Accelerator Laboratory,  Menlo Park, CA USA}
\affiliation{Department of Physics, Stanford University, Stanford, CA, USA}
 
\author{Jennifer Fowlie}
\affiliation{Stanford Institute for Materials and Energy Sciences, SLAC National Accelerator Laboratory,  Menlo Park, CA USA}
\affiliation{Department of Applied Physics, Stanford University, Stanford, CA, USA}

\author{Motoki Osada}
\affiliation{Stanford Institute for Materials and Energy Sciences, SLAC National Accelerator Laboratory,  Menlo Park, CA USA}
\affiliation{Department of Applied Physics, Stanford University, Stanford, CA, USA}

\author{Kyuho Lee}
\affiliation{Stanford Institute for Materials and Energy Sciences, SLAC National Accelerator Laboratory,  Menlo Park, CA USA}
\affiliation{Department of Physics, Stanford University, Stanford, CA, USA}

\author{Yonghun Lee}
\affiliation{Stanford Institute for Materials and Energy Sciences, SLAC National Accelerator Laboratory,  Menlo Park, CA USA}
\affiliation{Department of Applied Physics, Stanford University, Stanford, CA, USA}

\author{Danfeng Li}
\thanks{Present address: Department of Physics, City University of Hong Kong, Kowloon, Hong Kong}
\affiliation{Stanford Institute for Materials and Energy Sciences, SLAC National Accelerator Laboratory,  Menlo Park, CA USA}
\affiliation{Department of Applied Physics, Stanford University, Stanford, CA, USA}

\author{Harold Y. Hwang}
\affiliation{Stanford Institute for Materials and Energy Sciences, SLAC National Accelerator Laboratory,  Menlo Park, CA USA}
\affiliation{Department of Applied Physics, Stanford University, Stanford, CA, USA}

\date{\today}

\begin{abstract}
Infinite-layer nickelates present a new family of potential unconventional superconductors. A key open question is the superconducting pairing symmetry. We present low-temperature measurements of the London penetration depth in optimally doped La$_{0.8}$Sr$_{0.2}$NiO$_2$, Pr$_{0.8}$Sr$_{0.2}$NiO$_2$, and Nd$_{0.8}$Sr$_{0.2}$NiO$_2$. For La and Pr-nickelates, the superfluid density shows a quadratic temperature dependence, indicating nodal superconductivity in the presence of disorder. Nd-nickelate exhibits complex low-temperature behavior, which we attribute to magnetic impurities. These results are consistent with $d$-wave pairing. 
\end{abstract}

\maketitle
\listoffixmes
An important, enduring problem in condensed matter physics has been the nature and origin of superconductivity in the cuprates \cite{keimer2015}. In part due to this motivation, considerable effort has been directed toward searching for similar classes of superconductors  \cite{norman2016a}. A prominent example is the layered nickelates, due to nickel's proximity to copper in the periodic table \cite{lee2004,anisimov1999,lechermann2020a, botana2020} and shared structural and electronic aspects with cuprates. After extensive investigation, superconductivity was discovered in the infinite-layer nickelate (Nd,Sr)NiO$_2$ \cite{li2019a}. Initial examination of the electronic and magnetic properties has begun to reveal both similarities and differences between nickelates and cuprates \cite{hepting2021, li2020d, zeng2020}. The recent development of rare-earth variants of the infinite-layer nickelates, ($R$,Sr)NiO$_2$ for $R$=La or Pr, and subsequently quintuple-layer nickelates \cite{pan2021a}, has opened the door to exploration of a new family of superconducting compounds \cite{osada2020a, osada2021a, zeng2021a}. Furthermore, substantial improvements in crystallinity have made it possible to probe their properties at far greater depth than was previously possible \cite{lee2020a}. Here, we leverage these developments to investigate whether the superconducting gap has nodes, providing a `fingerprint' of the underlying pairing interaction \cite{hirschfeld2010}.

\begin{figure}
	\includegraphics[width=3.4 in]{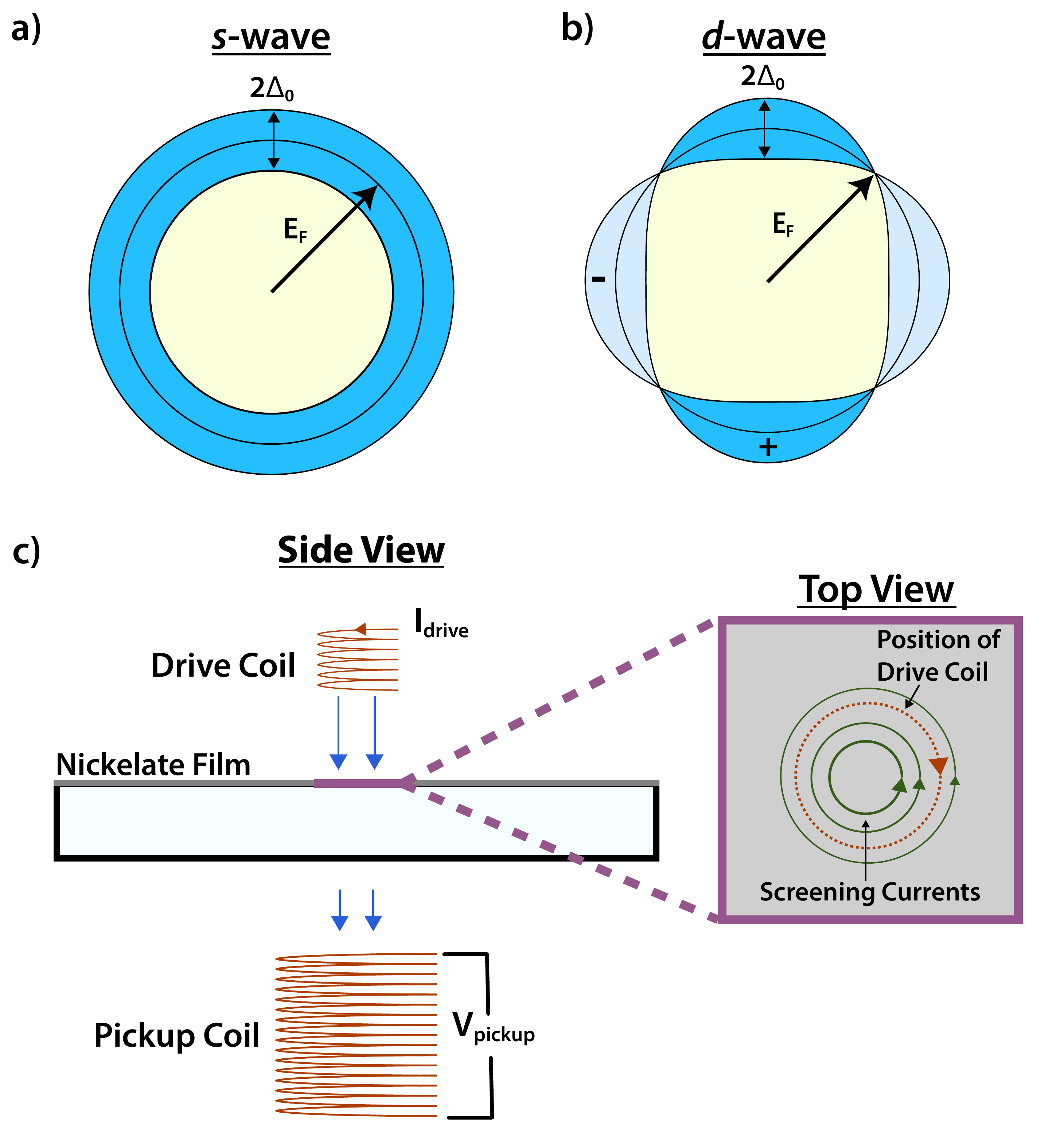}
	\caption{ a) A two-dimensional Fermi surface in momentum space for a conventional superconductor with isotropic $s$-wave pairing, which has a gap of constant magnitude $ \Delta_0 $ at the Fermi energy $E_F$.
	b) A two-dimensional Fermi surface for a $d$-wave superconductor possessing an anisotropic gap with nodes and sign changes in the gap. In particular, a gap with $d_{x^2-y^2}$ pairing defined by $ \Delta_0(\theta) = \Delta_0 \cos(2 \theta)$ is shown here.
	c) Schematic of the measurement apparatus. By sending a current $I_{\mathrm{drive}}$ through the drive coil directly above the nickelate sample, a magnetic field is generated at the sample (upper pair of blue arrows), inducing screening currents locally in the film, as shown in the Top View at right. The voltage across the pickup coil beneath the sample $V_{\mathrm{pickup}}$ measures the sum of the magnetic fields from the drive coil and the nickelate film (lower pair of blue arrows), allowing the film conductance and thereby the superfluid density to be calculated.}
	\label{fig1}
\end{figure}

The pairing interaction between electrons plays an important role in determining the wave function of the Cooper pairs and the corresponding superconducting gap. Electrons in conventional Bardeen-Cooper-Schrieffer (BCS) superconductors experience phonon-mediated attractive interactions and form an isotropic $s$-wave superconducting gap, as illustrated by \figref{fig1}{a} \cite{tinkham1996}. By contrast, cuprate superconductors form a $d$-wave superconducting gap (\figref{fig1}{b}) \cite{scalapino1995, tsuei2000}, which allows pairing in the presence of predominantly repulsive interactions between electrons. The pairing symmetry in nickelates is under active debate. Upper critical field measurements indicate spin-singlet pairing \cite{wang2021}. A number of theoretical studies have indicated a dominant $d_{x^2-y^2}$ pairing instability, but other mixed states have been proposed as well, in part due to the presence of electron pockets in the electronic structure in addition to the hole-like Fermi surface common with the cuprates \cite{nomura2019, wu2020a, zhang2020b, sakakibara2020, lechermann2020, werner2020}. A scanning tunneling spectroscopy experiment has observed spatially varying gap structures consistent with both $s$-wave and $d$-wave pairing \cite{gu2020, wu2020b}. The recent observation of strong antiferromagnetic spin fluctuations \cite{lu2021, krieger2021} could indicate favorable $d_{x^2-y^2}$ pairing. In this broad context, other experimental probes of the pairing symmetry are of strong interest.

To explore the superconducting gap structure, we perform measurements of the temperature dependence of the London penetration depth, the length scale over which magnetic fields decay in a superconductor, $\lambda(T)=\sqrt{\frac{m^*}{4 \mu_0 n_s(T) e^2}}$, where $m^*$ is the effective mass, $\mu_0$ is the vacuum permeability, $e$ is the electron charge, and $n_s$ is the superfluid density \cite{prozorov2006, hardy2002}. The change in superfluid density with temperature can be studied to determine whether the superconducting gap is nodal; at low temperatures $T\lesssim 0.3 T_c$, where $T_c$ is the superconducting transition temperature, changes in superfluid density are caused by thermal excitation of quasiparticles across the gap, which is a function of the shape and size of the superconducting gap as well as the temperature. An isotropic $s$-wave superconductor has a nonzero gap at all points, leading to an exponential decrease in superfluid density as temperature increases, so the normalized superfluid density follows the function

\begin{equation}
 \left(\lambda_0/\lambda(T)\right)^2= 1 - \sqrt{\frac{c_1}{\kb T}} \exp(-\Delta_0/\kb T),
 \label{eq:exp}
\end{equation} 

where $\lambda_0$ is the minimum value of the penetration depth, $c_1$ and $\Delta_0$ are fit parameters, and $\Delta_0$ represents the minimum gap size (Supplemental Information Sec. II  \footnote{See Supplemental Material for methods, details of fitting, and fits to the Arrhenius and log-log data sets.}).
By comparison, a gap with nodes, such as those with {\it d}-wave pairing, leads to a linear change in superfluid density with temperature, due to the nonzero quasiparticle density of states at the Fermi energy. When such superconductors have disorder, the linear behavior changes to quadratic at low temperatures due to the presence of impurity states at the Fermi energy. This can be parameterized using the equation 

\begin{equation}
    (\lambda_0/\lambda(T))^2 = 1 - 2 c_2 \frac{T^2}{T + T^{**}},
\label{eq:unit}
\end{equation}

where $c_2$ and $T^{**}$ are fit parameters and $T^{**}$ represents the crossover temperature between quadratic and linear behavior, with cleaner materials possessing lower values of $T^{**}/T_c$, and the cleanest cuprates reaching values of $T^{**}/T_c=0.01$ \cite{hirschfeld1993}. \fxnote{cite}

\begin{figure*}
	\includegraphics[width=6.45 in]{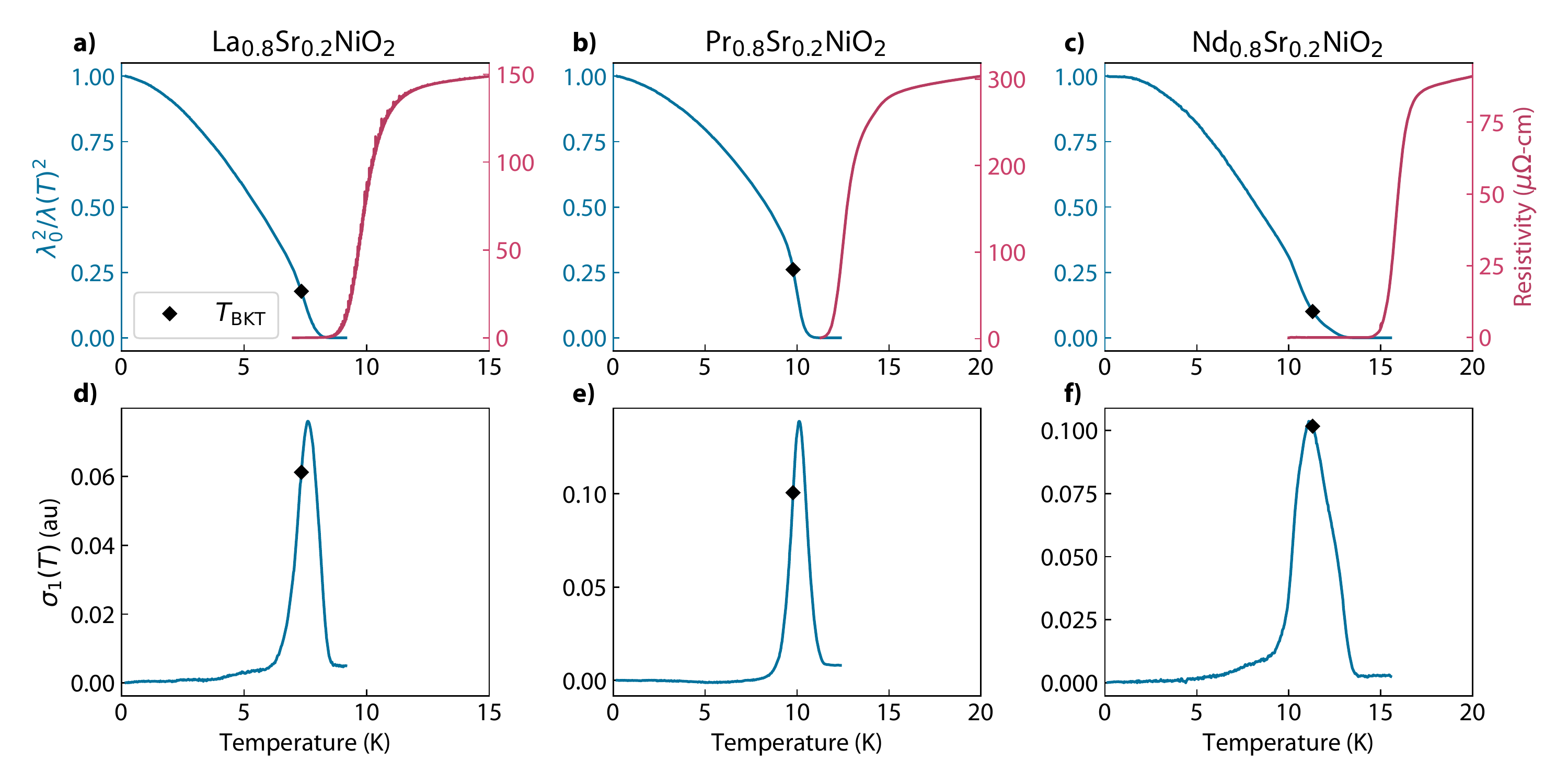}
	\caption{a-c) The normalized superfluid density $\lambda_0^2/\lambda(T)^2$ is plotted in blue (left axis) as a function of temperature for the three $R$-site variants of the infinite-layer nickelates. The resistivity of each sample is plotted in red (right axis), showing that the samples reach zero resistance at temperatures slightly above where the mutual inductance signal appears. d-f) $\sigma_1(T)$, the dissipative signal, extracted from the out-of-phase mutual inductance signal. A peak is seen at the superconducting transition, whose width reflects the homogeneity of the sample. All samples show a transition complete well-before $0.3 T_c$, so the low temperature behavior can be interpreted as representing the gap structure. The diamonds mark the BKT transition temperature $T_{\mathrm{BKT}}$.}
	\label{fig2}
\end{figure*}

\begin{figure*}[htp]
	\includegraphics[width=6.45 in]{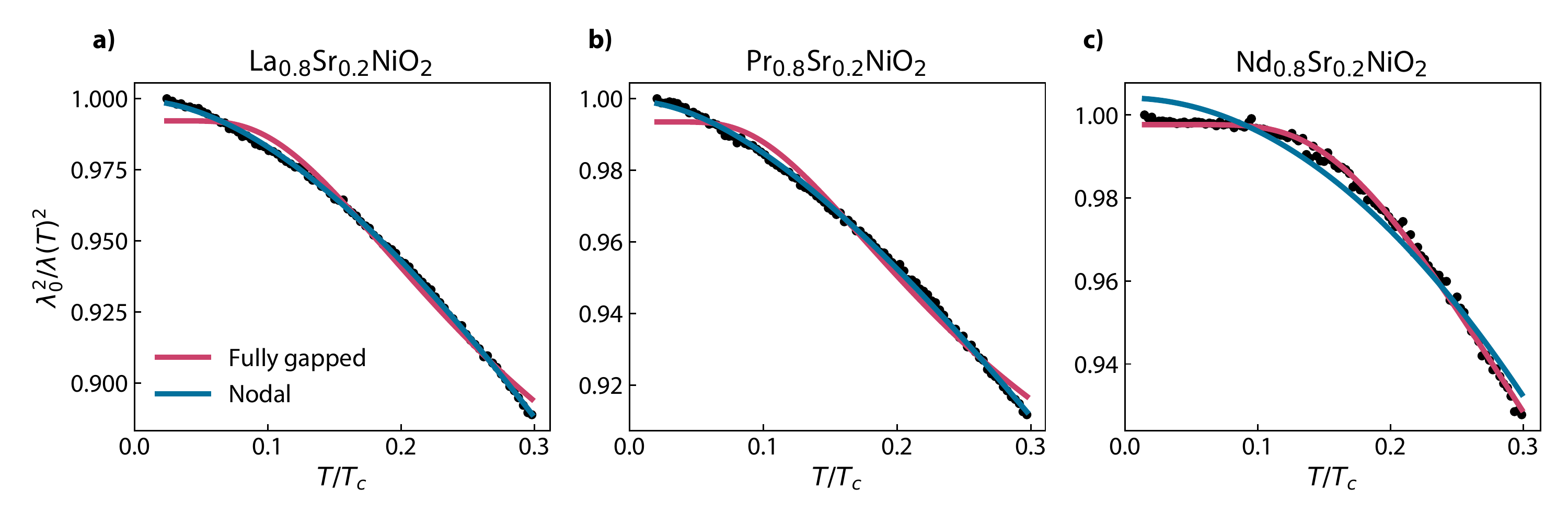}
	\caption{The normalized superfluid density plotted as a function of temperature below $0.3 T_c$ and fits to Eq. \eqref{eq:exp} (fully gapped) and Eq. \eqref{eq:unit} (nodal). a,b) Data for $R$=La,Pr show an excellent fit for the nodal equation. c) Data for $R$=Nd. The fit is superior for the fully gapped equation.}
	\label{fig3}
\end{figure*}

We measure the in-plane London penetration depth $\lambda_{ab}(T)$ (referred to from here on as $\lambda(T)$) using a mutual inductance two-coil technique optimal for thin films \cite{turneaure1998} in a dilution refrigerator (Supplemental Information, Sec. I). A diagram of our apparatus is shown in \figref{fig1}{c}; we send a current of 200-300 $\upmu$A at frequency $f=$ 30-60 kHz through the drive coil directly above the sample to generate a magnetic field at the sample, inducing screening currents in the film. The total magnetic field from the drive coil and screening currents is measured at a pickup coil directly below the sample. The complex conductance $\sigma_1(T) - i \sigma_2(T)$ can be extracted from the pickup voltage. $\sigma_2(T)$ represents the superconducting response, with superfluid density $n_s(T)= \frac{m^* \pi f}{ e^2} \sigma_2(T)$. $\sigma_1(T)$ represents the dissipative component of the signal; at the superconducting transition, we see a peak that gives a measure of the transition width and therefore the sample homogeneity \cite{bozovic2016}. $T_c$ throughout this manuscript is quantified as the temperature at which $\sigma_1(T)$ reaches its maximum. The sample is pressed into a sapphire plate for thermalization and a ruthenium oxide thermometer is attached to the sapphire to measure the sample temperature \cite{he2016}. Care is taken to determine that the sample remains in the linear response regime.

\begin{table*}[htp]
\begin{center}
\begin{tabular}{|c|c|c|c|c|c|c|c|c|} 
 \hline
 R & $T_c$ (K) & $\lambda_0$ ($\upmu$m) &$c_1$ & $\Delta_0 $ & $c_2 $ & $T^{**} $ & $c_2'$ \\ 
 La & 7.6    & 1.35  & 0.09 (0.04)  & 0.5 (0.1) & 0.5 (0.12)  & 0.5 (0.18) & - \\ 
 Pr & 10.1   & 1.3   & 0.04  (0.02) & 0.5 (0.1) & 0.3 (0.06) & 0.3 (0.12)  & - \\
 Nd & 11.1   & 0.75  & 0.3 (0.1)    & 0.8 (0.07) & -          & -          & 0.4 (0.025) \\
 \hline
\end{tabular}
\end{center}
\caption{\label{tab:table1} Properties and fit parameters of the samples discussed. $T_c$ here is the value at which $\sigma_1$ is maximum. $\lambda_0$ is the minimum penetration depth. $c_1$ and $\Delta_0$ are fit parameters for Eq. 1, and both are normalized by $k_{\mathrm{B}} T_c$ in the table. $c_2$ and $T^{**}$ are fit parameters for Eq. 2, both normalized by $T_c$ in the table. For Nd where $T^{**} \gg T_c$, a quadratic fit with coefficient $c_2' = c_2/T^{**}$ is appropriate and is presented instead. Errors in fit parameters are displayed in parentheses.} 
\end{table*}

We present measurements of each of the $R$-site variants $R$=La, Pr and Nd at optimal doping $R\mathrm{_{0.8}Sr_{0.2}NiO_2}$. Samples are approximately 7 nm thick and further capped with SrTiO$_3$, grown using pulsed laser deposition of the perovskite form of the material on a SrTiO$_3$ substrate followed by soft chemical reduction to the infinite-layer phase, as described elsewhere \cite{lee2020a}. Careful optimization of the film growth has improved film quality considerably, resulting in lower resistivity and superconducting transition width as well as lower defect density in scanning transmission electron microscopy \cite{lee2020a, osada2021a}. The resistivity and normalized superfluid density $\lambda_0^2/\lambda(T)^2$ for the three samples is plotted in \figref{fig2}{a-c}, showing that the resistance reaches zero at temperatures 1-2 K higher than where the superfluid density appears. The real conductance is plotted in \figref{fig2}{b}, showing that the  transition width measured through mutual inductance is 1-2 K, and for all samples, the transition is completed at a temperature well over $0.3 T_c$, which is necessary for the low temperature behavior to be a reliable guide to the pairing symmetry of the material. A Berezinskii-Kosterlitz-Thouless (BKT) transition, exhibiting an abrupt change in superfluid density close to $T_c$, can be seen for each sample, indicating that they are highly homogeneous \cite{maccari2017}. The temperature at which this is predicted to happen, $T_{\mathrm{BKT}}$, is defined by the point at which $8 \pi \mu_0 / \phi_0^2 \kb T_{\mathrm{BKT}}  = d/\lambda(T_{\mathrm{BKT}})^2$, where $d$ is the sample thickness and $\phi_0$ is the flux quantum. This point is marked by a diamond in each panel of Figure 2, from which we see that it is located slightly below the temperature for the onset of superfluid density. This is due to disorder in the material and is often observed in superconducting thin films \cite{yong2013}.  $\lambda_0$ varies from 0.75 $\upmu$m to 1.35 $\upmu$m. These values are substantially larger than predicted by density-functional-theory calculations \cite{bernardini2020}, most likely as a result of disorder.

\begin{figure*}
	\includegraphics[width=6.45 in]{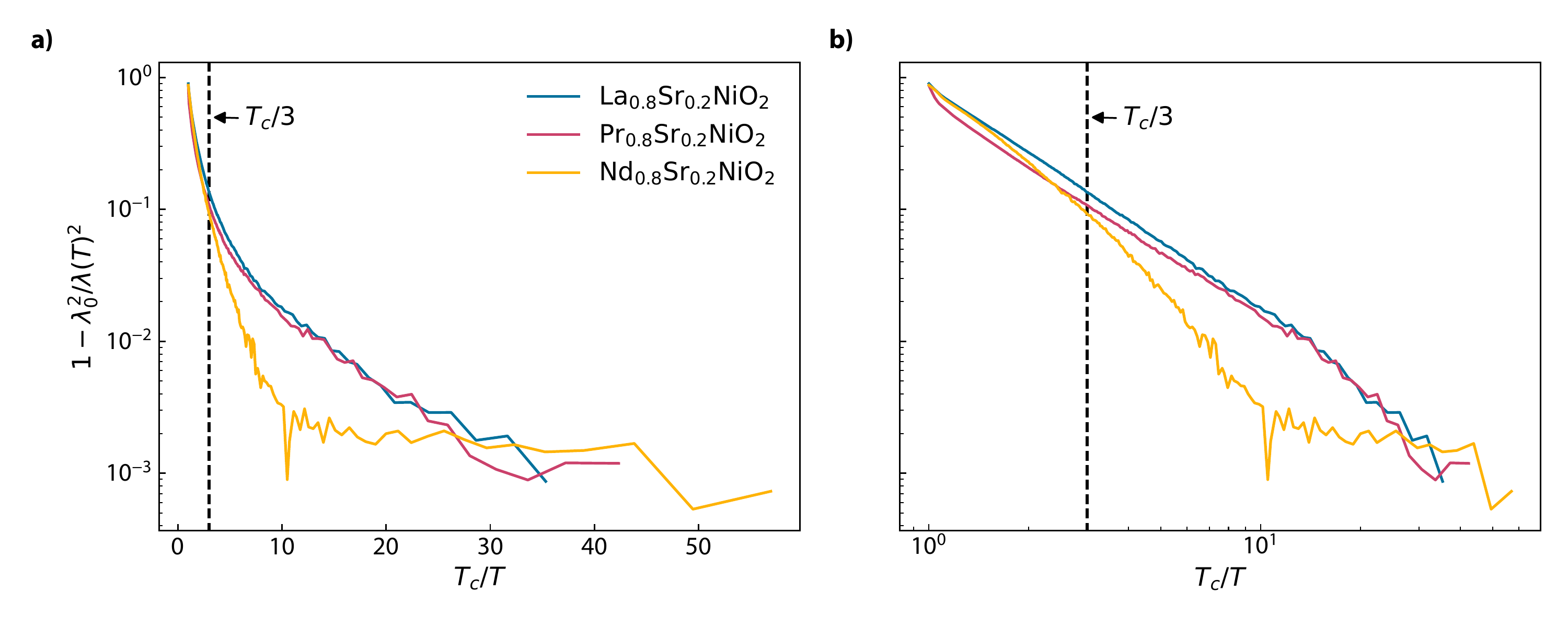}
	\caption{a) Arrhenius plot of the normalized inverse temperature versus the normalized change in superfluid density. For a fully gapped state, a straight line below $T_c/3$ (dashed line) is expected. The La and Pr samples do not exhibit this behavior anywhere, while Nd shows a straight line between approximately $T_c/3$ and $T_c/10$ ($T\approx 1$ K) before abruptly flattening at lower temperatures. b) Log-log plot of the normalized inverse temperature versus the normalized change in superfluid density. Power-law scaling appears as a straight line here, as observed for the La and Pr samples from near $T_c$ down to the noise floor, with slopes of 1.7 (between linear and quadratic scaling). The Nd samples exhibit a straight line from near $T_c$ down to 3-4 K, after which the behavior becomes steeper and then flatter.}
	\label{fig4}
\end{figure*}

In \figref{fig3}{}, the low temperature behavior of the samples and fits to Eqs. \ref{eq:exp} and \ref{eq:unit} are shown, with the fit parameters displayed in Table \ref{tab:table1}. For La and Pr, we find that the data is fit closely by Eq. \ref{eq:unit} and poorly by Eq. \ref{eq:exp}, indicating that the gap is nodal and consistent with {\it d}-wave pairing. For both samples, $T^{**}$ is of the order of $T_c$, consistent with materials with an intermediate level of disorder. These materials are in the dirty limit of superconductivity, with the electron mean free path shorter than the superconducting coherence length \cite{wang2021}, so this is expected. We present an Arrhenius plot and log-log plot of the change in the normalized superfluid density against normalized inverse temperature for all the samples measured in \figref{fig4}{a-b}. Surprisingly, the La and Pr samples, exhibit a straight line in \figref{fig4}{b} below about 85\% $T_c$ down to almost the lowest temperatures, indicating power-law scaling, with slope of 1.7 (Supplemental Material, Sec. III). While this is expected at low temperatures given the excellent fit to Eq. \eqref{eq:unit}, it is unexpected that it would continue to such a high temperature. This may have similar origin to the linear scaling of superfluid density with temperature seen in cuprate superconductors \cite{broun2007, lee-hone2017}.

The apparent superfluid density of Nd, shown in \figref{fig3}{c}, flattens as the temperature is reduced below 1 K. Because mutual inductance directly measures the magnetic field generated by the sample, it is sensitive to magnetic impurities in the material, which have been seen to impact penetration depth measurements in other superconductors containing rare-earth elements such as Nd \cite{cooper1996, prozorov2000a, martin2009}. This can impact the measured value of the penetration depth $\lambda_{\mathrm{meas}}(T)=\sqrt{\mu(T)}\lambda(T)$ where $\mu(T)$ is the magnetic permeability of the material. We also note that the Nd sample has the broadest and irregular transition (\figref{fig1}{f}). Therefore, while the best fit for Nd is the exponential rather than the quadratic function, this is unlikely to result from the superconducting gap being nodeless, and we ascribe this to the presence of magnetic defects. Another indication that the state is not nodeless comes from the fit parameters of the exponential, which suggest the minimum gap size is equal to $0.8 \kb T_c$, well below the BCS weak-coupling limit of 1.76$T_c$. Finally, examination of the Arrhenius plot for Nd-nickelate sample presented in \figref{fig4}{a} shows that it does not follow exponential scaling; while a straight line is expected, we instead see a straight line at higher temperatures, followed by an abrupt flattening, evidence that the flattening at 1 K does not follow an exponential dependence.

In summary, we have performed measurements of the penetration depth in the infinite-layer nickelates down to 150 mK, and found that La$_{0.8}$Sr$_{0.2}$NiO$_2$ and Pr$_{0.8}$Sr$_{0.2}$NiO$_2$ exhibit quadratic scaling in the low temperature regime, consistent with a superconductor with impurities and a nodal gap, and consistent with dirty $d$-wave superconductivity. By contrast, Nd$_{0.8}$Sr$_{0.2}$NiO$_2$ displays complex behavior inconsistent with simple models for nodal and fully gapped superconductivity, which is most likely to be a function of magnetic impurities. These results situate the infinite-layer nickelates as unconventional superconductors, and likely analogs to high-temperature cuprates in their pairing symmetry.

\textit{Note:} During the preparation of this manuscript, we became aware of another study of the penetration depth in nickelates \cite{chow2022}. A similar low-temperature magnetic contribution was observed for Nd$_{0.8}$Sr$_{0.2}$NiO$_2$. Other aspects of the data, analysis and symmetry conclusion are different from our findings.

\acknowledgements{We thank Doug Bonn, Varun Harbola, Peter Hirschfeld, Aharon Kapitulnik, Steve Kivelson,  Peter Littlewood, Sri Raghu, and David Saykin for useful discussions. This work was supported by the US Department of Energy, Office of Basic Energy Sciences, Division of Materials Sciences and Engineering (contract no. DE-AC02-76SF00515) and the Gordon and Betty Moore Foundation’s Emergent Phenomena in Quantum Systems Initiative (grant no. GBMF9072, synthesis equipment and development of the mutual inductance probe). J.F. was also supported by the Swiss National Science Foundation through Postdoc.Mobility P400P2199297 and Division II 200020 179155.}

\bibliography{bib}
\end{document}


\title{Supplemental Information \\
Evidence for nodal superconductivity in infinite-layer nickelates}

\author{Shannon P. Harvey}
\thanks{These authors contributed equally}%
\affiliation{Stanford Institute for Materials and Energy Sciences, SLAC National Accelerator Laboratory,  Menlo Park, CA USA}
\affiliation{Department of Applied Physics, Stanford University, Stanford, CA, USA}

\author{Bai Yang Wang}
\thanks{These authors contributed equally}
\affiliation{Stanford Institute for Materials and Energy Sciences, SLAC National Accelerator Laboratory,  Menlo Park, CA USA}
\affiliation{Department of Physics, Stanford University, Stanford, CA, USA}

\author{Jennifer Fowlie}
\affiliation{Stanford Institute for Materials and Energy Sciences, SLAC National Accelerator Laboratory,  Menlo Park, CA USA}
\affiliation{Department of Applied Physics, Stanford University, Stanford, CA, USA}

\author{Motoki Osada}
\affiliation{Stanford Institute for Materials and Energy Sciences, SLAC National Accelerator Laboratory,  Menlo Park, CA USA}
\affiliation{Department of Applied Physics, Stanford University, Stanford, CA, USA}

\author{Kyuho Lee}
\affiliation{Stanford Institute for Materials and Energy Sciences, SLAC National Accelerator Laboratory,  Menlo Park, CA USA}
\affiliation{Department of Physics, Stanford University, Stanford, CA, USA}

\author{Yonghun Lee}
\affiliation{Stanford Institute for Materials and Energy Sciences, SLAC National Accelerator Laboratory,  Menlo Park, CA USA}
\affiliation{Department of Applied Physics, Stanford University, Stanford, CA, USA}

\author{Danfeng Li}
\thanks{Present address: Department of Physics, City University of Hong Kong, Kowloon, Hong Kong}
\affiliation{Stanford Institute for Materials and Energy Sciences, SLAC National Accelerator Laboratory,  Menlo Park, CA USA}
\affiliation{Department of Applied Physics, Stanford University, Stanford, CA, USA}

\author{Harold Y. Hwang}
\affiliation{Stanford Institute for Materials and Energy Sciences, SLAC National Accelerator Laboratory,  Menlo Park, CA USA}
\affiliation{Department of Applied Physics, Stanford University, Stanford, CA, USA}

 \date{\today}

\maketitle
\section{Methods}
\subsection{Mutual Inductance Set Up} 
Coils were formed using 20 $\upmu$m diameter copper wire to limit the total coil diameter to under 0.5 mm to reduce the impact of edge currents. The pickup coil is composed of 400 turns and has dimensions of approximately 0.5 mm x 0.5 mm. The drive coil is composed of 50 turns and has dimensions 0.25 mm x 0.25 mm. Leakage around the sample was measured at 1.5\% using a 100 nm thick aluminum film with lateral dimensions equal to those of the sample (2.5 mm x 5 mm). At this film thickness, there will be no transmission through the film itself, so the mutual inductance measured reflects leakage around the film. Measurements of nickelate samples are calibrated for this leakage by subtraction.

\begin{figure}
	\includegraphics[width=6.4 in]{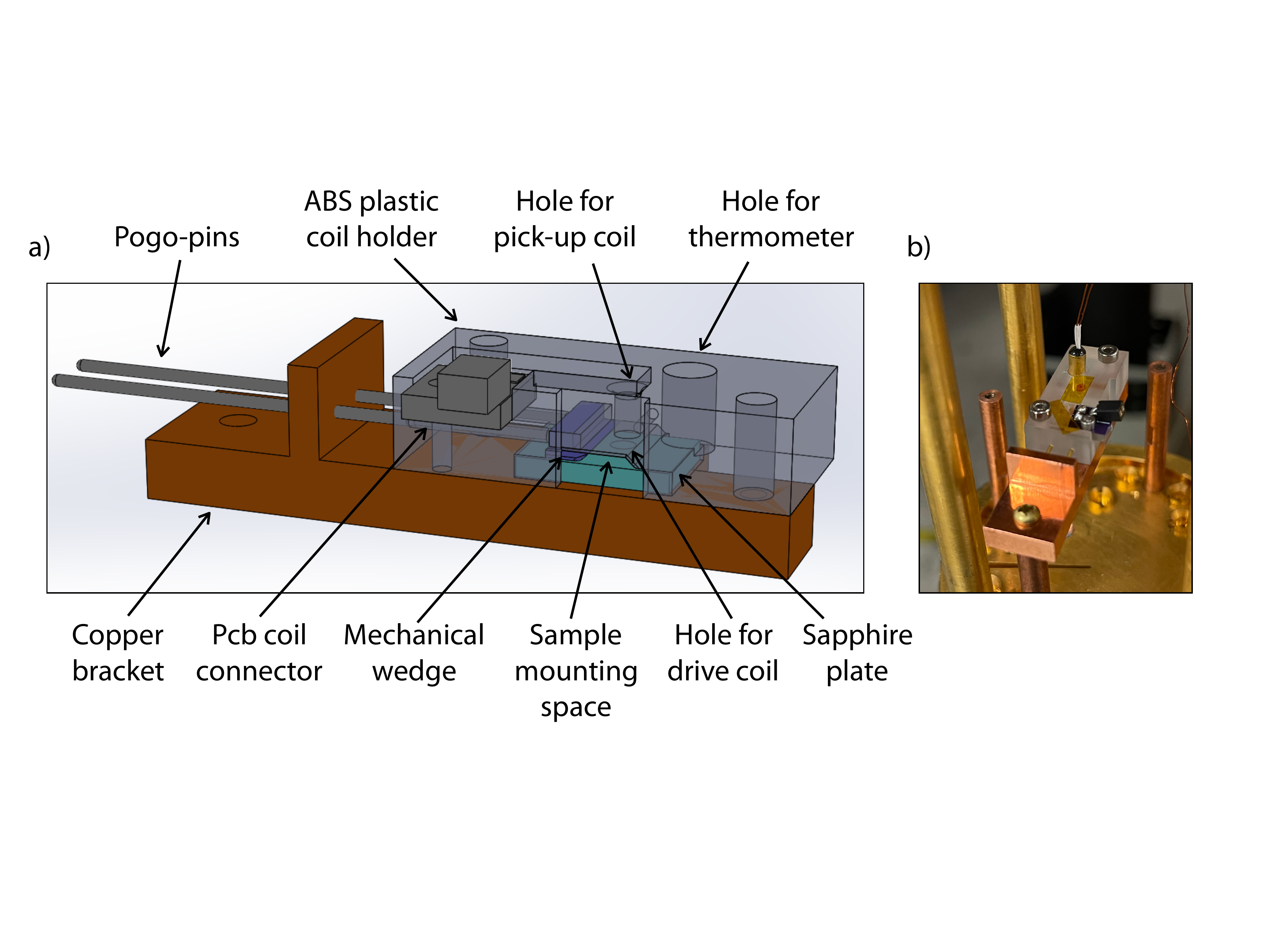}
	\caption{Schematic of the set up for mutual inductance measurements. a) Drawing showing the key facets of the device. The device follows the design shown in \cite{he2016}, with pogo pins pressing a mechanical wedge into the sample, which thereby presses it into a sapphire plate for thermalization. The coils are positioned directly above and below the sample, and the entire set up is aligned by a copper bracket, which connects it to the mixing chamber. A RuO$_2$ thermometer is glued with silver paint to the sapphire plate as well to provide a local measurement of temperature. b) Photograph of the actual device, affixed to the dilution refrigerator.}  
\label{schem}
\end{figure}

\subsection{Measurements}
Measurements were performed with a lock-in amplifier, during recovery of the helium mixture, while the temperature was increased at between 2-5 mK / minute in the temperature range below $T_c/3$ and 10-25 mK/min at the temperature range above. To reduce noise, data is binned in 30 mK increments. 

We measure a voltage, $V_{\mathrm{pickup}}= 2 \pi f I_{\mathrm{drive}} M$, where $M$ is the mutual inductance between the two coils. For an infinitely-wide sample, this can be solved 
\begin{align}
    M(\lambda, \sigma_1) &=\Sigma_{i} \Sigma_{j} \pi \mu_0 r_{\mathrm{dr}}^i r_{\mathrm{pu}}^j \int_0^{\infty} dx \frac{e^{- x D^{i,j}} J_1 (x r_{\mathrm{dr}}^i) J_1(x r_{\mathrm{pu}}^j)}{\cosh(d \chi) + \frac{x^2 + \chi^2}{2 \chi x} \sinh(d \chi)} \\
    \chi^2 &= x^2 + 1/\lambda^2 + 2 i \pi \mu_0 f \sigma_1 \nonumber, 
\end{align}

where $r_{\mathrm{dr}}^i $ is the radius of the $i^\mathrm{th}$ turn of the drive coil, $r_{\mathrm{pu}}^j $ is the radius of the $j^\mathrm{th}$ turn of the pickup coil, $d$ is the sample thickness, $D^{i,j}$ is the distance between the $i^\mathrm{th}$ turn of the drive coil and the $j^\mathrm{th}$ turn of the pickup coil. The integral term is summed across each combination of a turn of the drive coil and turn of the pickup coil. We calculate this term for a series of penetration depths and for $\lambda=\infty, \sigma_1=0$, which we use to normalize the signal $M_{\mathrm{norm}} = M_x(\lambda, 0)/M_x(\infty, 0)$, thereby reducing the error from uncertainty in geometry. We estimate the error in the absolute value of the penetration depth is about 10$\%$ using this method, dominated by uncertainty in the coil geometry. The error in the temperature dependence of the penetration depth is far lower.  

The phase of the measurement is set by auto-phasing the signal at base temperature for our first sample measurement, and keeping the phase constant for the remaining measurements. The phase of the measurement is within 5 degrees of $90^{\circ}$, with the difference most likely occurring due to capacitances in the wiring.  The signal can be represented as $ M = M_x + i M_y$. We extract the penetration depth by normalizing $M_x$ by the signal measured directly above $T_c$, $M_{x,n}$, and looking up the result by interpolating a table of $M_{\mathrm{norm}}(\lambda,0)$, calculated as above.

For subsequent measurements, we correct the $M_y$ signal at low temperatures by subtracting an offset from it to set it to 0 at base temperature, which gives the most physical results for the temperature dependence of $M_y(T)$, with $M_y(T)$ never becoming negative. We expect that the small changes in the signal magnitude of about $0.02 M_x$ are due to parasitics in the lines. In the limit where $\sigma_1$ is small, which is true in our experiment, $M_y \propto \sigma_1$, and values of $\sigma_1$ plotted in this paper are equal to $M_y/M_{x,n}$. Due to uncertainty about the exact impact of the wiring, these results are not quantitative, but act as a qualitative guide to the superconducting transition.
\subsection{Transport Measurements} 
The samples were contacted using wire-bonded aluminum wires and measured in a Quantum Design Physical Property Measurement System with base temperature of 2 K. 
\section{Fit details}
 
The fit for the nodal gap is derived in \cite{hirschfeld1993}. For a conventional BCS $s$-wave gap that is constant for all values in $k$-space (such as the one shown in Figure 1a), it is straightforward to derive the temperature dependence of the superfluid density, 
 \begin{equation}
 \left(\lambda_0/\lambda(T)\right)^2= 1 - \sqrt{\frac{2 \pi \Delta_0}{\kb T}} \exp(-\Delta_0/\kb T) 
 \end{equation}
 This provides a poor fit to all of the data sets, including the Nd-nickelate. Because the nodal equation (Eq. 2 of main text) has two fit variables, while this equation has only one, this fit is more constrained than the nodal one. Therefore, we use a variant of the BCS equation that has a second fit variable (Eq. 1 of main text) to test for a non-nodal gap. This might be explained physically, for instance, by a multiple gaps or a gap that varies in size while remaining non-nodal.  

\section{Fit to data in Arrhenius and log-log plots}
Fits are shown for each of the Arrhenius and power law fits, and the slopes are presented in Table S1. The slope of the Arrhenius plot represents the magnitude of the superconducting gap, $\Delta_0/\kb T_c$. The weak-coupling limit for $s$-wave pairing is $1.76 T_c$. That the slopes of the Arrhenius plot for the Nd-nickelate is far below that is a strong indication that they do not possess conventional $s$-wave pairing. Some pnictide materials possessing $s^{\pm}$ pairing have been measured with gaps of similar magnitude. The La- and Pr-nickelate samples indicate gaps of under $0.2T_c$, suggesting that if there is a non-nodal gap, it has an extremely small minimum. 

The slope of the log-log plot, $m$, represents the power-law scaling of the penetration depth, $1-\lambda_0^2/\lambda(T)^2=c T^m$ where $c$ is an arbitrary constant. The slopes of the data sets from the La- and Pr-nickelates are 1.7. If the temperature range from $0.85T_c$ to 3 K is used for Nd-nickelate, the range where magnetism is unlikely to contribute, we find that the data set fits a power law with scaling of 2. However, the significance of this not clear, because this temperature range is mostly above $T_c/3$ and may therefore reflect the temperature dependence of the gap rather than the pairing symmetry of the material. 

\begin{figure}
	\includegraphics[width=6.4 in]{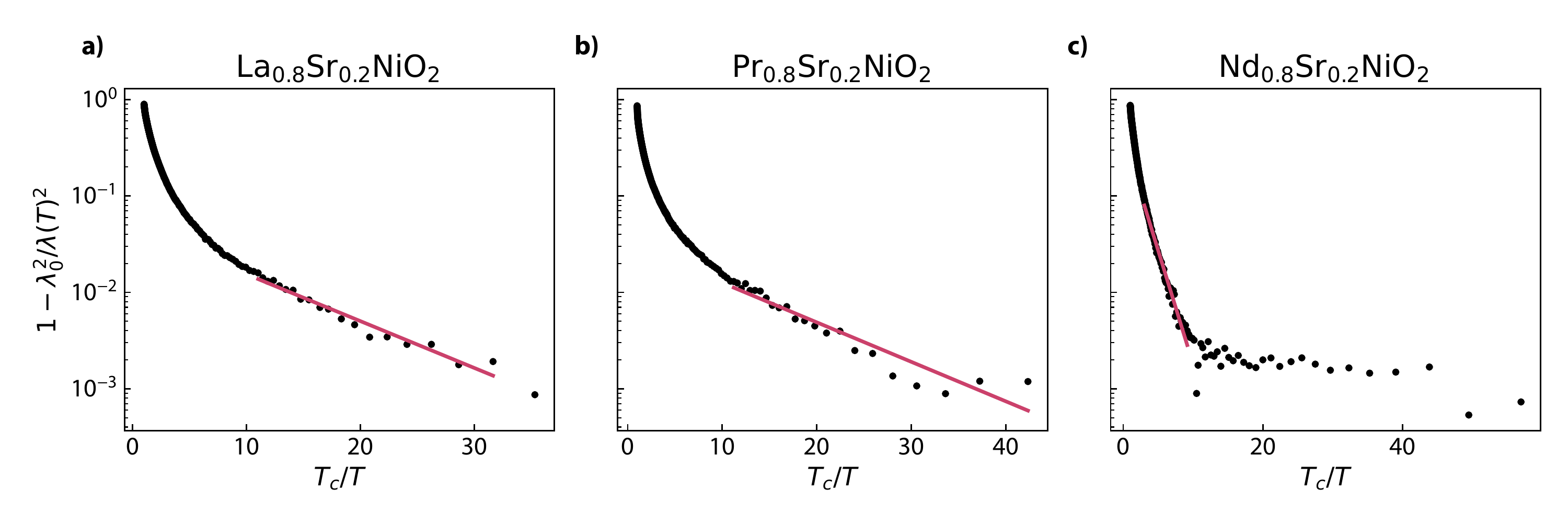}
	\caption{Arrhenius plots for all samples. For La-nickelate and Pr-nickelate, the line is a fit to the low temperature data ($T/T_c<0.1$). Nd-nickelate shows distinctly different behavior at low temperatures, most likely due to magnetic effects, so the data is fit to a line between $T/T_c<0.3$ and $T>1.2$ K. The slopes are presented in Table S1.}  
\label{arr}
\end{figure}

\begin{figure}
	\includegraphics[width=6.4 in]{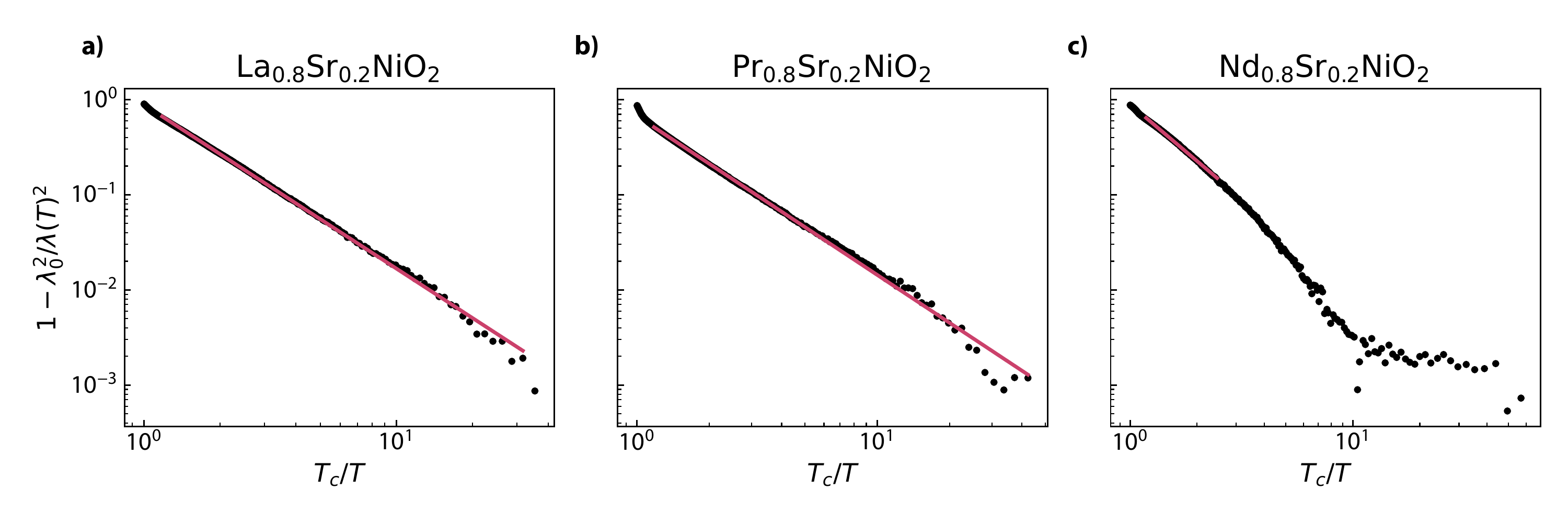}
	\caption{Log-log plots for all samples, labeled with the sample number. For La-nickelate and Pr-nickelate, the line is fit from $0.85T_c$ down to base temperature. The fit-line extends across the straight-line region of the Nd-nickelate, which extends from $0.85 T_c$ to about 3 K. The slopes are presented in Table S1.}
\label{pow}
\end{figure}

\begin{table}
\begin{center}
\begin{tabular}{|c|c|c|c|c|c|c|c|c|} 
 \hline
 R & $T_c$ (K) & Exp. slope & Pow. Slope \\ 
 La & 7.6    & 0.1 (0.005)  & 1.7 (0.03)  \\ 
 Pr & 10.1   & 0.1  (0.08) & 1.7 (0.04)  \\
 Nd & 11.1   & 0.5 (0.015)  & 2 (0.02) \\
 \hline
\end{tabular}
\end{center}
\caption{\label{tab:table1} Slopes of the Arrhenius and Log-log plots, listed as Exp. slope and pow. slope, respectively. Errors in fit parameters are displayed in parentheses.}
\end{table}

\bibliography{bib}